\documentclass[twocolumn,amsmath,amssymb]{revtex4}

\usepackage{graphicx}
\usepackage{bm}

\begin{document}

\title{A rotating cavity for high-field angle-dependent microwave spectroscopy of low-dimensional conductors and magnets}

\author{Susumu Takahashi}
\affiliation{Department of Physics, University of Florida,
Gainesville, FL 32611, USA}
\author{Stephen Hill}
\affiliation{Department of Physics, University of Florida,
Gainesville, FL 32611, USA}

\date{\today}

\begin{abstract}
The cavity perturbation technique is an extremely powerful method
for measuring the electrodynamic response of a material in the
millimeter- and sub-millimeter spectral range (10~GHz to 1~THz),
particularly in the case of high-field/frequency magnetic resonance
spectroscopy. However, the application of such techniques within the
limited space of a high-field magnet presents significant technical
challenges. We describe a $7.62$~mm~$\times~7.62$~mm
(diameter~$\times$~length) rotating cylindrical cavity which
overcomes these problems. The cylinder is mounted transverse to the
bore of the magnet, coupling is achieved through the side walls of
the cavity, and the end plate is then rotated (by means of an
external drive) instead of the body of the cavity itself. Therefore,
rotation does not affect the cylindrical geometry, or the mechanical
connections to the incoming waveguides. The TE011 mode frequency of
the loaded cavity is 51.863~GHz, with the possibility to work on
higher-order modes to frequencies of order 350~GHz. Neither the
quality factor ($\sim22,000$ for the fundamental mode) or the
coupling to the cavity are significantly affected for full
$360^\circ$ of rotation. The rotation mechanism provides excellent
angle resolution ($<0.1^\circ$), and is compact enough to enable
measurements in the high-field (up to 45~T) magnets at the National
High Magnetic Field Laboratory. Two-axis rotation capabilities are
also possible in conjunction with split-pair magnet configurations.
We present examples of angle-dependent measurements which illustrate
the unique capabilities of this rotating cavity, including:
high-field angle-dependent measurements of a novel form of cyclotron
resonance in anisotropic organic conductors; and angle-dependent
high-frequency single-crystal electron paramagnetic resonance (EPR)
measurements in single-molecule magnets.
\end{abstract}


\maketitle

\section{\label{sec:level1}INTRODUCTION}
In recent years, microwave (millimeter and sub-millimeter wave)
technologies, covering frequencies from 10~GHz to 10~THz
($0.33-330$~cm$^{-1}$), have become the focus of intensive efforts
in many fields of research. In engineering and medicine, THz imaging
represents one of the next-generation technologies, enabling
non-destructive materials inspection, chemical composition
analysis~\cite{Mittleman96:tray,Mittleman99:timage,herrmann02:timag},
and medical
diagnoses~\cite{Mittleman96:tray,Mittleman99:timage,han00:timage}.
In the fundamental sciences, physics, chemistry and biology,
microwave spectroscopy is also very useful for investigating the
physical properties of a material. This is particularly true for the
sub-field of condensed matter physics, where the millimeter and
sub-millimeter spectral range can provide extremely rich information
concerning the basic electronic characteristics of a
material~\cite{klein93, donovan93, dressel93, Bonn}. Furthermore,
combining microwave techniques and high magnetic fields (microwave
magneto-optics), allows many more possibilities, including:
cyclotron resonance
(CR)~\cite{kip61:cr,moore62:cr,hill97,blundell97:por,ardavan98:khg,hill00,kovalev02,kovalev03:prl};
electron paramagnetic resonance
(EPR)~\cite{Poole,hill98:epr,hill03:sci,hill03:prl};
antiferromagnetic resonance (AFMR)~\cite{Kittel}; Josephson Plasma
Resonance (JPR) measurements of layered
superconductors~\cite{matsuda95,gaifullin97,mola00JPR,hill02:jpr};
and many others. In each of these examples, the magnetic field
influences the dynamics of electrons at frequencies spanning the
microwave spectral range. Another very important aspect of microwave
magneto-optical investigations is the possibility to study angle
dependent effects by controlling the angle between the sample and
the microwave and DC electromagnetic fields. For instance, through
studies of the angle dependence of CR amplitudes, one can extract
detailed information concerning the Fermi surface (FS) topology of a
conductor~\cite{kovalev02,kovalev03:prl}. Consequently,
angle-dependent microwave spectroscopy has been widely used in
recent years to study highly anisotropic magnetic and conducting
materials. Problems which have been addressed using these methods
include: high-Tc superconductivity~\cite{matsuda95}, and other
low-dimensional superconductors, e.g. organic
conductors~\cite{hill97,blundell97:por,ardavan98:khg,kovalev02,kovalev03:prl,mola00JPR,hill02:jpr},
Sr$_{2}$RuO$_{4}$~\cite{hill00}, etc.;
the quantum and fractional
quantum Hall effects~\cite{engel93:qhe,li97:qhe}; and
low-dimensional magnets, including single-molecule magnets
(SMMs)~\cite{hill98:epr,hill03:sci,hill03:prl}.

Unfortunately, the microwave spectral range presents many technical
challenges, particularly when trying to study very tiny ($\ll
1$~mm$^3$) single-crystal samples within the restricted space inside
the bore of a large high-field magnet system (either resistive or
superconducting~\cite{magnetspec}). Problems associated with the
propagation system stem from standing waves and/or
losses~\cite{mola00}. Several methods have been well documented for
alleviating some of these issues, including the use of fundamental
TE and TM mode rectangular metallic waveguides, low-loss cylindrical
corrugated HE waveguides~\cite{smith98}, quasi-optical propagation
systems~\cite{smith98}, and in-situ generation and detection of the
microwaves~\cite{TDS}. Standing waves are particularly problematic
in the case of broadband spectroscopies, e.g. time-domain~\cite{TDS}
and fourier transform techniques~\cite{FTIR}, as well as for
frequency sweepable monochromatic sources~\cite{goy98:spie,BWO}. In
these instances, the optical properties are usually deduced via
reflectivity or transmission measurements, requiring a large
well-defined (i.e. flat) sample surface area ($>\lambda^2$). For
cases in which large samples are not available (note: $\lambda$
spans from 3~cm at 10~GHz to 0.3~mm at 1~THz), resonant techniques
become necessary, e.g. cavity
perturbation~\cite{klein93,donovan93,dressel93,Poole,mola00}. This
unfortunately limits measurements to the modes of the cavity. In
addition, making absolute measurements of the optical constants of a
sample, as a function of frequency, is extremely difficult to
achieve using cavity perturbation because of its narrow-band
nature~\cite{klein93,donovan93,dressel93}. However, the cavity
perturbation technique is ideally suited for fixed-frequency,
magnetic resonance measurements~\cite{mola00}. Furthermore, as we
have recently shown, it is possible to make measurements at many
different frequencies by working on higher order modes of the
cavity~\cite{mola00}. To date, measurements in enclosed cylindrical
copper cavities have been possible at frequencies up to 350~GHz (see
Sec.~\ref{smm}).

The standard approach for studying angle-dependent effects using the
cavity perturbation technique is to use a split-pair magnet and/or
goniometers, and is widely used in lower frequency commercial EPR
instruments, e.g. X-band, K-band and
Q-band~\cite{datars61:sb,kip61:cr}. In the case of the split-pair
approach, the DC magnetic field is rotated with respect to a static
waveguide/cavity assembly. In this paper, we outline a method for
in-situ rotation of part of a cylindrical resonator, thus enabling
angle-dependent cavity perturbation measurements in ultra-high-field
magnets, and two-axis rotation capabilities in standard high-field
superconducting split-pair magnets. As we shall outline, the
rotation mechanism preserves the cylindrical symmetry of the
measurement, thereby ensuring that the electromagnetic coupling to
the microwave fields does not change upon rotating the sample. This
is particularly important for studies of low-dimensional conductors,
where sample rotation alone (as in the case of a goniometer) would
lead to unwanted instrumental artifacts associated with
incommensurate symmetries of the sample and cavity. The available
frequency range in the setup developed at the University of Florida
(UF) spans from 8~GHz up to $\sim 700$~GHz (0.27-23~cm$^{-1}$).
Experiments are possible in DC magnetic fields of up to 17 tesla,
and at temperatures in the range from $0.5-400$~kelvin at UF.
Furthermore, all experimental probes and techniques developed at UF,
including the ones described in this article, are also compatible
with the 45~tesla resistive magnets at the National High Magnetic
Field Laboratory (NHMFL) in Tallahassee, FL.

We note that a rotating cavity has previously been developed by
Schrama et al.~\cite{schrama01}, at the University of Oxford, also
for high-field microwave studies. As we will demonstrate in this
article, the cylindrical geometry offers many advantages over the
rectangular design implemented by the Oxford group. For example, the
waveguides are coupled rigidly to the cylindrical body of the cavity
in our design (only the end-plate rotates), whereas the coupling is
varied upon rotation in the Oxford version, resulting in effective
``blind-spots", i.e. angles where the microwave fields in the
waveguides do not couple to the cavity; in contrast, the cylindrical
version offers full $360^\circ+$ rotation. Furthermore, the rigid
design offers greater mechanical stability and, therefore, less
microwave leakage from the cavity, resulting in improved
signal-to-noise characteristics~\cite{mola00}. The TE01$n$
cylindrical modes also offer the advantage that no AC currents flow
between the curved surfaces and flat rotating end-plate of the
cylindrical resonator. Consequently, the moving part of the cavity
does not compromise the exceptionally high quality ($Q-$) factors
associated with these modes. Indeed, $Q-$factors for the first few
TE$01n$ modes vary from $10,000$ to $25,000$ (at low temperatures),
as opposed to just $500$ for the rectangular cavities. This order of
magnitude improvement translates into vastly increased sensitivity,
enabling e.g., EPR studies of extremely small single crystal
samples. Finally, the cylindrical cavity is machined entirely from
copper, and held rigidly together entirely by screws. This
all-copper construction results in negligible field sensitivity,
i.e. the field dependence of the cavity parameters is essentially
flat and, most importantly, the cavities do not contain any
paramagnetic impurities that could give rise to spurious magnetic
resonance signals.

In the next section, the detailed features of the rotating cavity
are described, and some example data obtained with the rotating
cavity will be shown in the following section.

\begin{table*}
\caption{\label{tab:table1}~Available magnet systems at the University of Florida (UF) and the NHMFL. The table lists the field geometry and magnet type, the
maximum available field $B_{max}$, the probe length, the available temperature (T) range, and the outer diameter of the cavity probe.}
\begin{ruledtabular}
\begin{tabular}{llllll}
Magnet & $B_{max}$~(T) & Type & length & T (kelvin) & Probe dia.\\
\hline
45~T (NHMFL) & 45 & Axial hybrid & 1.67 m & 1.4 - 300 & 3/4"\\
33~T (NHMFL) & 33 & Axial resistive & 1.45 m & 1.4 - 300 & 3/4"\\
25~T (NHMFL) & 25 & Axial resistive & 1.6 m & 0.5 - 300\footnote{~$0.5-1.4$~K with a $^3$He refrigerator.} & 1" \\
Oxford Inst. (UF) & 17 & Axial SC & 1.9 m & 0.5 - $300^a$ & 1" (7/8"$^a$)\\
QD PPMS (UF) & 7 & Transverse split-coil SC\footnote{~Allows two-axis rotation in the Quantum Design (QD) PPMS system~\cite{PPMS}.}   & 1.15 m & 1.7 - 400 & 1"\\
\end{tabular}
\end{ruledtabular}
\end{table*}

\section{\label{sec:level2}Technical Description}
\subsection{\label{sec:level21}Overview of experimental setup}
As a tunable source of millimeter and sub-millimeter-wave radiation,
we use a Millimeter-wave Vector Network Analyzer (MVNA), including
an External Source Association (ESA) option. The basic
MVNA-8-350-1-2 analyzer utilizes pairs of phase-locked sweepable
($8-18.5$~GHz) YIG sources and multipliers (Schottky diodes),
enabling superheterodyne vector (phase sensitive) measurements, with
sufficient dynamic-range for the measurements described in this
paper, to frequencies on the order of 250~GHz. Further details
concerning the MVNA are published in refs.~\cite{mola00,goy98:spie}.
The ESA option additionally permits association with a higher
frequency Gunn diode source, providing enhanced signal-to-noise
characteristics in the $150-250$~GHz range, as well as enabling
measurements to considerably higher frequencies (up to $700$~GHz, or
$ 23$~cm$^{-1}$~\cite{goy98:spie}). For the purposes of this
article, we note that the ESA option can be used very effectively
for studies on higher order modes of the rotating cavity ($150$ up
to almost $400$~GHz $\--$ see section~III). A detailed description
of the use of the MVNA for cavity perturbation measurements up to
$\sim200$~GHz, including a description of the waveguide coupling
between the MVNA and the low-temperature/high-field environment, has
been presented previously by Mola et al.~\cite{mola00}. For
measurements above 400~GHz, we employ a quasi-optical bridge coupled
to a simple reflectivity probe (with no cavity) constructed from
low-loss cylindrical HE mode waveguide; this instrumentation will be
described elsewhere~\cite{Edwards04}.

The various magnet systems which are compatible with the instrumentation described in this article are listed in TABLE~\ref{tab:table1}. The standard cryostats
designed for these magnets are all $^4$He based (either bath or flow cryostats). Because of the demand to work at lower temperatures, we have constructed a
simple $^3$He refrigerator which is compatible with the 17~tesla Oxford Instruments superconducting magnet at UF, and the 25~T resistive magnets at the NHMFL,
as listed in TABLE~\ref{tab:table1}. A schematic of this refrigerator is shown in Fig.~1.
\begin{figure}
\includegraphics[width=50 mm]{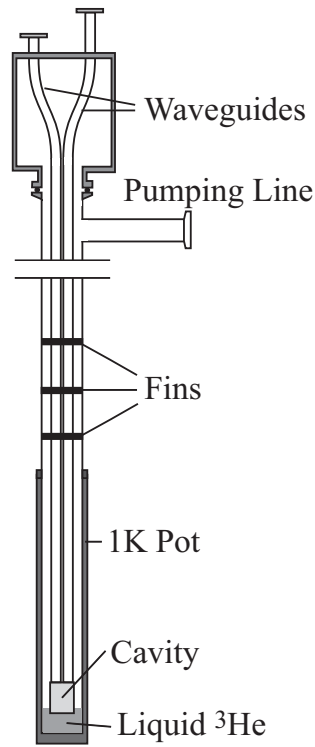}
\caption{\label{fig:he3probe} A schematic diagram of the $^3$He probe used for sub-kelvin experiments in the Oxford Instruments 17~T superconducting magnet.
See main text for a detailed description of its construction.}
\end{figure}
The 1.9~m long waveguide/cavity probe is inserted directly into the $^3$He space, which is constructed from a 7/8" ($=22.2$~mm) outer diameter stainless steel
tube with a $0.010"$ ($=0.25$~mm) wall thickness. The lower 254~mm of this tube is double jacketed with a $1.00"$ ($=25.4$~mm) outer diameter. The volume
between the two tubes is vacuum sealed in order to provide thermal isolation between the $^3$He liquid and the surrounding $^4$He vapor. The $^3$He condenses
by means of heat exchange with the walls of the 7/8" tube (above the double jacketed region) which is inserted into the Oxford Instruments $^4$He flow cryostat
operating at its base temperature of $\sim1.4$~K. After condensation of the full charge of $^3$He (5~liters at STP), sub-kelvin temperatures are achieved by
pumping directly on the liquid by means of an external sealed rotary pump. The refrigerator operates in single-shot mode, i.e. the $^3$He is returned to a room
temperature vessel, where it is stored until the next cooling cycle. A simple gas handling system controls the condensation of $^3$He gas, and the subsequent
pumping of the gas back to the storage vessel. The $^3$He tube and gas handling system is checked for leaks prior to each cool down from room temperature.
Although this design is simple, it has the disadvantage that the microwave probe comes into direct contact with the $^3$He vapor, thus potentially affecting
the tuning of the cavity, as well as the phase of the microwaves reaching the cavity via over 3.8~m of waveguide; such phase fluctuations can cause drifts in
signal intensity due to unavoidable standing waves in the waveguide. However, we have found that these problems are minimal when operating at the base $^3$He
vapor pressure (0.15 torr). The temperature of the sample is then controlled by supplying heat to the copper cavity, which acts as an excellent heat reservoir,
i.e. it ensures good thermal stability. The base temperature of the $^3$He refrigerator is 500 mK and it provides hold times of up to 2 hours.

\subsection{\label{sec:level22}Rotating cavity}
The configuration of the rotating cavity is shown in Fig.~2.
\begin{figure}
\includegraphics[width=70 mm]{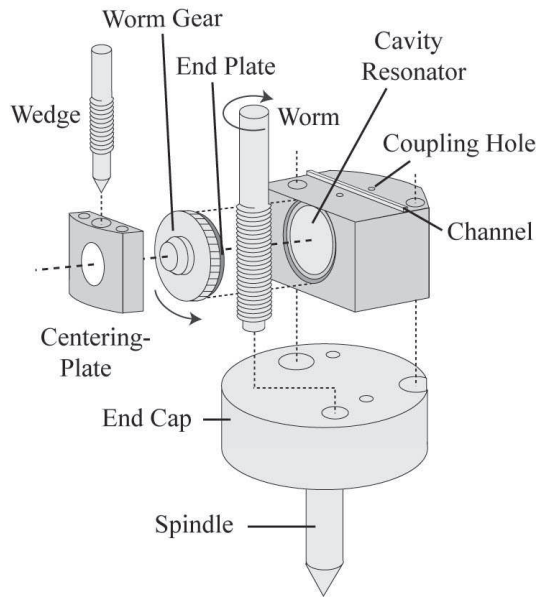}
\caption{\label{fig:rcavity} A schematic diagram of the various components that make up the rotating cavity system. The sample may be placed on the end plate,
which can then be rotated via an externally controlled worm drive. The wedge is used to clamp and un-clamp the end plate to/from the main resonator. See main
text for a detailed description of the assembly and operation of the rotating cavity.}
\end{figure}
The principal components consist of the open-ended cylindrical resonator, the cavity end-plate and worm gear, a worm drive for turning the end-plate, and a
wedge which facilitates external clamping and un-clamping of the cavity and end-plate. The cavity assembly is mounted on the under-side of a stage (not shown
in Fig.~2), and the end-plate is centered on the axis of the cavity by means of a centering-plate. The upper part of the wedge is threaded, and passes through
a threaded channel in the stage so that its vertical position can be finely controlled via rotation from above. Likewise, the worm drive is rotated from above,
and accurately aligned with the worm gear via an unthreaded channel in the stage. Finally, the worm-drive and centering plate are additionally constrained
laterally by means of an end cap and spindle (see below) which attaches to the under side of the resonator. The cavity, end plate, and stage are each machined
from copper, thus ensuring excellent thermal stability of the environment surrounding the sample; the heater and thermometry are permanently contacted directly
to the stage. The remaining components shown in Fig.~2 are made from brass~\cite{note1}.

The internal diameter of the cavity (7.62~mm) is slightly less than the diameter of the end-plate, which is free to rotate within a small recess machined into
the opening of the resonator. On its rear side, the copper end-plate mates with a brass gear which, in turn, rotates on an axis which is fixed by the
centering-plate. As mentioned above, rotation of the worm gear and end-plate is achieved by turning the worm drive with the wedge disengaged from the
end-plate. During experiments, a good reproducible contact between the end-plate and the main body of the cavity is essential for attaining the highest
resonance $Q$-factors. This is achieved by engaging the wedge through a vertical channel in the centering-plate, where it transfers pressure along the axis of
the end-plate. Fig.~3 displays labeled photographs of the 1st and 2nd generation rotating cavity assemblies.
\begin{figure}
\includegraphics[width=50 mm]{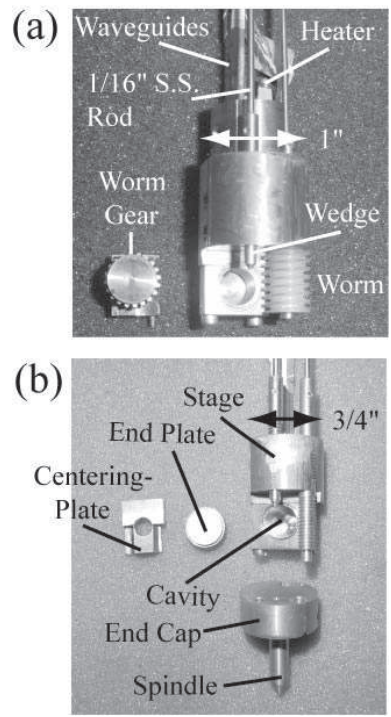}
\caption{\label{fig:rotatorpic} Photographs of the rotating cavity system; part of the cavity has been disassembled (end-plate and centering-plate) in order to
view the inside of the cylindrical resonator. (a) shows the 1st generation rotating cavity, which fits into a 1" thin-walled stainless steel tube; this cavity
is compatible with the 25~T magnets at the NHMFL and the QD PPMS and Oxford Instrument magnets at UF (see Table~\ref{tab:table1}). (b) shows the 2nd generation
rotating cavity, which fits into a 3/4" thin-walled tube; this cavity is compatible with the highest field (45~T) resistive magnets at the NHMFL, as well as
the $^3$He probe designed for the Oxford Instruments magnet at UF (see Fig.~1).}
\end{figure}
The 2nd generation version employs a smaller home-built worm drive, reducing the overall diameter of the probe to slightly below 3/4" ($=19.1$~mm), which
enables its use in the highest field magnets at the NHMFL; the 1st generation probe has an outer diameter of just under 1".

The wedge and worm gear are driven by stainless steel rods
(Diameter~=~1/16", or 1.59~mm) which pass through vacuum tight
`o'-ring seals at the top of the waveguide probe. Small set screws
are used to fix the steel control rods into the worm drive and
wedge (Fig.~2), and to fix the end-plate within the worm gear.
Rotation of the worm gear is monitored via a simple turn-counting
dial mounted at the top of the probe, having a readout resolution
of 1/100th of a turn. Different worm drive/gear combinations are
employed in the 3/4" and 1" diameter probes (see Fig.~3), with
1/41 and 1/20 gear ratios, respectively. Thus, the angle
resolution on the dial redout corresponding to the actual sample
orientation is approximately $0.09^\circ$ for the 3/4" probe, and
$0.18^\circ$ for the 1" probe. Although both probes exhibit
considerable backlash ($\sim 1^\circ$), this is easily avoided by
consistently varying the sample orientation in either a clockwise
or counter-clockwise sense. High resolution EPR measurements on
single-molecule magnets (reported in section~III) have confirmed
the angle resolution figures stated above.

The stage also performs the task of clamping the V-band waveguides
into position directly above the cavity coupling holes. As with
previous cavity designs, a small channel [0.02" ($=0.51$~mm) wide
and 0.02" deep] is machined between the waveguides on the under
side of the stage; this channel mates with a similarly sized ridge
located in between the coupling holes on the upper surface of the
cavity housing (see Fig.~2). Our previous studies have
demonstrated that this arrangement is extremely effective at
minimizing any direct microwave leak between the incident and
transmission waveguides and is, therefore, incorporated into all
of our cavity designs~\cite{mola00}. A direct leak signal can be
extremely detrimental to cavity perturbation measurements, causing
a significant reduction in the useful dynamic range, and to
uncontrollable phase and amplitude mixing, as explained in our
previous paper~\cite{mola00}.

The internal dimensions of the cylindrical resonator are $7.62$~mm~$\times~7.62$~mm (diameter $\times$ length). The center frequency of the TE011 mode of the
unloaded cavity ($f_\circ$) is 51.863~GHz, with the possibility to work on higher-order modes as well. Using the ESA option, we have been able to conduct
measurements up to 350~GHz~\cite{Edwards04}. While the modes above about 150~GHz are not well characterized, they do provide many of the advantages of the
well-defined lower frequency modes, e.g. enhanced sensitivity, control over the electromagnetic environment (i.e. E vs. H field) at the location of the sample,
and some immunity to standing waves. Table~\ref{tab:resonances} shows resonance parameters for several unloaded cavity modes (there are many others which are
not listed). The TE$01n$ ($n=$ positive integer) modes are probably the most important for the rotating cavity design, because their symmetry is axial. Thus,
rotation of the end-plate not only preserves the cylindrical symmetry, but also ensures that the sample remains in exactly the same electromagnetic field
environment, i.e. upon rotation, the polarization remains in a fixed geometry relative to the crystal. A sample is typically placed in one of two positions
within the cavity: i) directly on the end-plate; and ii) suspended along the axis of the cavity by means of a quartz pillar (diameter = 0.75~mm) which is
mounted in a small hole drilled into the center of the end-plate. These geometries are depicted in Fig.~4 for the TE011 mode, and for the two DC magnetic field
geometries, i.e. axial and transverse.
\begin{figure}
\includegraphics[width=50 mm]{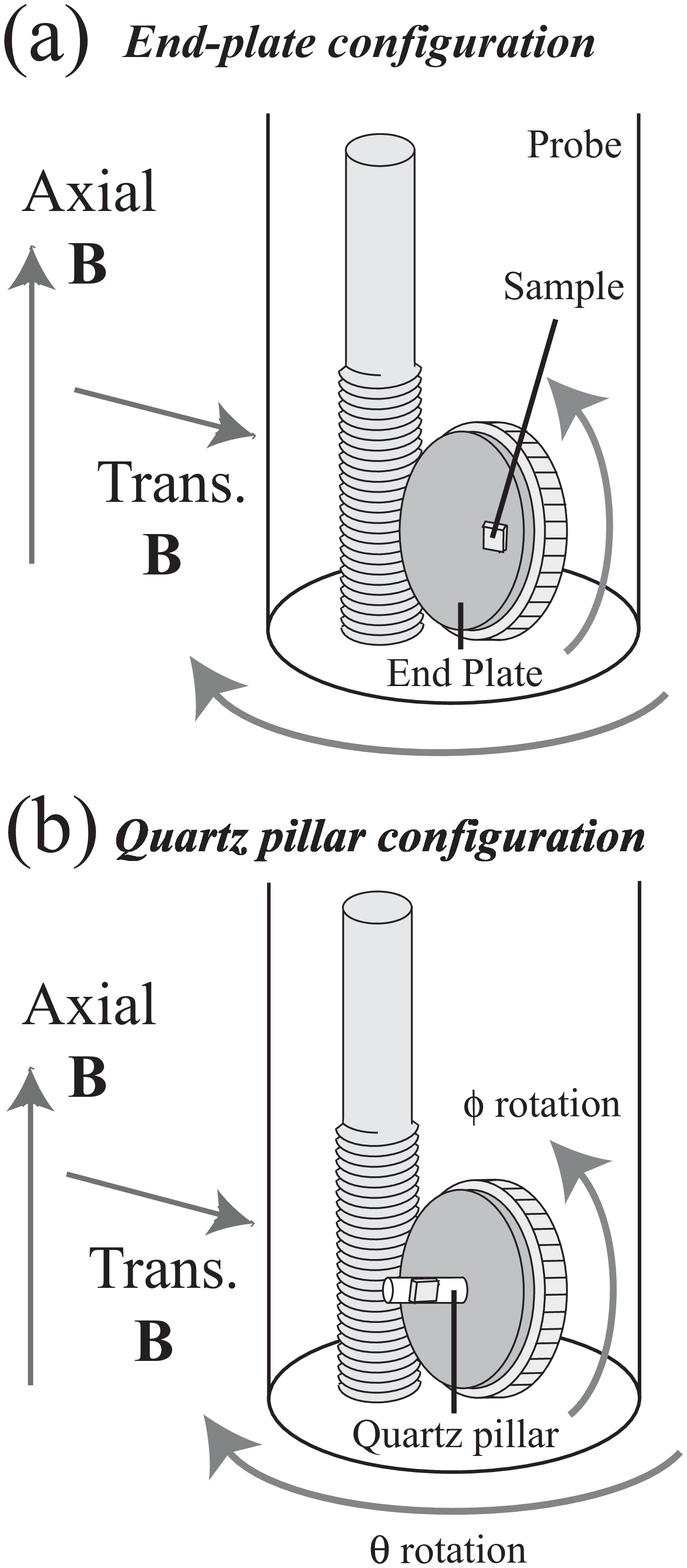}
\caption{\label{fig:rotation} Schematic diagrams showing various different sample mounting configurations for both axial and transverse magnetic field
geometries, including the two-axis rotation capabilities.}
\end{figure}
Each geometry possesses certain advantages for a particular type of experiment. For example, the quartz pillar configuration is particularly useful for EPR
experiments in the axial high-field magnets [Fig.~4(b)] since, for the TE$01n$ modes, the sample sits in a microwave AC field ($\tilde{H}_1$) which is always
transverse to the DC magnetic field ({\bf B}$_\circ$). We discuss this in more detail in section~III, and in ref.~\cite{mola00}.

Another advantage of the TE$01n$ modes is the fact that no microwave
currents flow between the end-plate and the main body of the
resonator. Thus, the $Q$-factors of these modes are high, and
essentially insensitive to the mechanical contact made with the
wedge. Table~\ref{tab:resonances} lists the key resonance parameters
associated with several modes.
\begin{table}
\caption{\label{tab:resonances}Resonance parameters for several different cavity modes. The first column indicates the given mode. The second column lists the
resonance frequencies ($f_\circ$). The third column lists the $Q$-factors. The final column lists the contrast in dB, i.e. A($f_\circ)-$A$_l$, where
A($f_\circ$) is the transmission amplitude at the resonance frequency, $f_\circ$, and A$_l$ is the leak amplitude.}
\begin{ruledtabular}
\begin{tabular}{cccc}
Mode & $f_0$(GHz) & $Q$ & A($f_0$)-A$_l$(dB) \\
\hline
TE011 & 51.863 & 21,600 & 31.7 \\
TE212 & 54.774 & 3,300 & 26.5 \\
TE012 & 62.030 & 16,600 & 25.2 \\
TE015 & 109.035 & 8,600 & 19.0 \\
\end{tabular}
\end{ruledtabular}
\end{table}
For the case of the TE011 mode, the
low-temperature ($\sim2$~K) $Q$-factor is $\sim21,600$, and the
contrast between the amplitude on resonance [A($f_\circ$)] and the
amplitude far from resonance (leak amplitude, A$_l$) is 31.7dB.
These parameters are essentially the same as the optimum values
reported for the fixed cylindrical geometry in our earlier
paper~\cite{mola00}, thus confirming the suitability of this new
rotating design for cavity perturbation studies of small
single-crystal samples, both insulating and conducting. We note that
the $Q$-value for the non-cylindrically symmetric mode in
Table~\ref{tab:resonances} (TE212) is almost an order of magnitude
lower than that of the TE011 mode. As noted above, this is due to
the flow of microwave currents associated with the TE212 mode across
the mechanical connection between the main body of the resonator and
the end-plate.

As discussed earlier, even though the end-plate can be rotated,
the waveguides are coupled absolutely rigidly to the cavity via
the stage. As with earlier designs~\cite{mola00}, the microwave
fields in the waveguides are coupled into the resonator by means
of small circular coupling holes which are drilled through the
sidewalls of the cavity. The sizes of these coupling holes
[diameter = 0.038" (=0.97~mm), or $\sim\lambda/6$] have been
optimized for the V-band, and the cavity side-wall was milled down
to a thickness of 0.015" (=0.38~mm, or $\sim\lambda/15$) at the
location of these holes. Once again, these numbers are essentially
the same as those reported in ref.~\cite{mola00}. The key point
here is that this coupling never changes during rotation.
Therefore, the full 360$^\circ$ angle range may be explored, and
with excellent mechanical stability. Fig.~5 shows the typical
random fluctuations in the cavity resonance parameters for the
TE$01n$ mode during a complete $360^\circ$ rotation of the
end-plate: the center frequency varies by no more than
$\pm120$~kHz ($\sim3$~ppm, or $5\%$ of the resonance width); the
$Q$-factor is essentially constant, to within $\pm1\%$; and the
contrast, [A($f_\circ)-$A$_l$], fluctuates between 29~dB to 33~dB,
corresponding respectively to leak amplitudes of $3.5\%$ and
$2.2\%$ of A($f_\circ)$.
\begin{figure}
\includegraphics[width=70 mm]{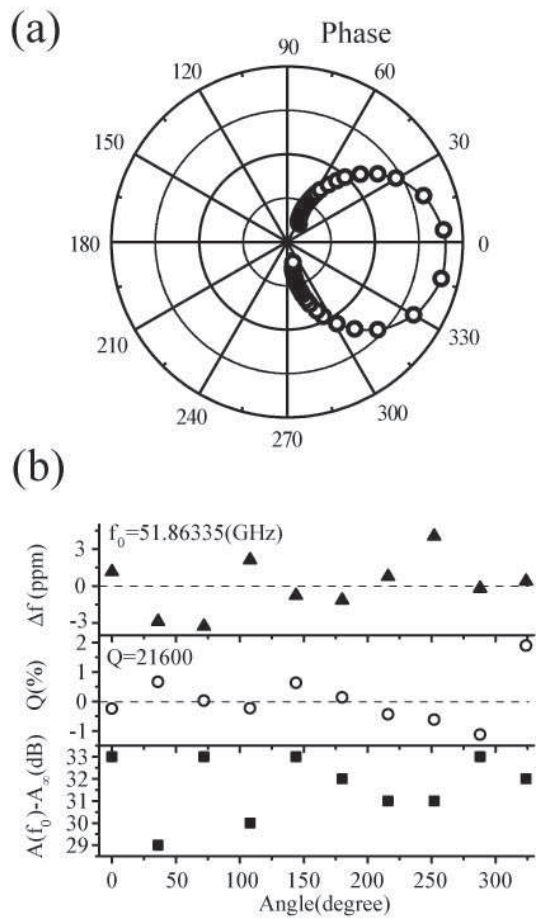}
\caption{\label{fig:te011} (a)~A polar plot of the complex signal transmitted through the cavity as the frequency is swept through the TE011 mode; the polar
coordinate represents the phase of the transmitted signal, while the radial coordinate corresponds to the linear amplitude. The Lorentzian resonance is
observed as a circle in the complex plane. Points close to the origin correspond to frequencies far from resonance, while the resonance frequency corresponds
to the point on the circle furthest from the origin. The resonance parameters are obtained from fits (solid line) to the data (open circles) in the complex
plane (see ref.~\cite{mola00} for a detailed explanation of this procedure). The average center frequency, $f_\circ$, is 51.863~GHz, and the $Q$ factor is
21600. (b)~Full $360^\circ$ angle-dependence of the fluctuations in the resonance parameters for the TE011 mode: upper panel $\--$ $f_\circ$; center panel
$\--$ $Q$; and  lower panel $\--$ difference between the amplitude on resonance [$A(f_\circ$)] and the amplitude far from resonance (the leak amplitude,
$A_l$). These data illustrate the excellent reproducibility of the resonance parameters upon un-clamping, rotating, and re-clamping the cavity.}
\end{figure}
Another important consequence of coupling through the side walls of the cavity involves selection rules for TM cylindrical cavity modes. At the lowest
frequencies, only the TE01 mode of the V-band waveguide propagates. Thus, the microwave $\tilde{H}_1$ field in the waveguide is polarized parallel to the
cavity axis, i.e. it is incompatible with the symmetry associated with the $\tilde{H}_1$ patterns of the TM modes. Consequently, we do not observe, for
example, the TM111 mode. This offers added benefit, since the TM$11n$ modes are ordinarily degenerate with the TE$01n$ modes, and steps have to be taken to
either lift these degeneracies, or to suppress the TM modes all together~\cite{mola00}. Here, we simply do not couple to these modes at the lowest frequencies.
Even at the highest frequencies, we do not expect coupling to TM modes, provided the polarization of the microwave sources is maintained throughout the
waveguide.

The main design challenge in the development of a useful rotating
cavity system concerned the space constraints imposed by the
high-field magnets at the NHMFL; the specifications of each magnet
system are listed in Table~\ref{tab:table1}. A prototype
configuration was first developed, based on a $1"$ outer tube
diameter. This prototype was subsequently implemented in both
magnet systems at UF, and in the 25~T, 50~mm bore resistive
magnets at the NHMFL. A picture of this cavity system, which
remains in use, is shown in Fig.~3(a). The major reason for the
large size of the cavity assembly is the large worm drive diameter
($3/8"=9.525$~mm), which is determined by the smallest readily
available commercial components~\cite{sdpsi}; this 1st generation
worm drive is made of nylon. A more compact 2nd generation
rotating cavity was developed by machining a considerably smaller
custom worm drive (and gear) in-house. This cavity assembly, which
is shown in Fig.~3(b), is small enough to fit into a 3/4" diameter
thin-walled stainless steel tube, thus enabling measurements in
the highest field 45~T hybrid magnet at the NHMFL. Furthermore,
this cavity is compatible with the $^3$He probe constructed for
the 17~T Oxford Instruments superconducting magnet (and the 25~T
resistive magnet at the NHMFL), allowing for experiments at
temperatures down to 500~mK. We note that the 1st and 2nd
generation cavity assemblies may be transferred relatively easily
from one particular waveguide probe to another, requiring only
that the thermometry be unglued and re-glued to the stage using GE
varnish; all other connections are made with screws. In addition,
we have constructed extra parts for both designs, including
several cavities and end-plates. This enables preparation of a new
experiment while an existing experiment is in progress.

Finally, we discuss the two-axis rotation capabilities made
available via a 7~T Quantum Design (QD) magnet (see
Table~\ref{tab:table1}). The 7~T transverse QD system is outfitted
with a rotation stage at the neck of the dewar. A collar clamped
around the top of the waveguide probe mounts onto this rotation
stage when the probe is inserted into the PPMS
flow-cryostat~\cite{PPMS}. The rotator is driven by a computer
controlled stepper motor, with $0.01^\circ$ angle resolution. The
motor control has the advantage that it can be automated and,
therefore, programmed to perform measurements at many angles over an
extended period of time without supervision. In the $50-250$~GHz
range, the compact Schottky diodes can be used for microwave
generation and detection. These devices are mounted directly to the
probe, and are linked to the MVNA via flexible coaxial cables
(feeding the diodes with a signal in the $8-18$~GHz range).
Therefore, the waveguide probe can rotate with the source and
detector rigidly connected, while the vacuum integrity of the flow
cryostat is maintained via two sliding `o'-ring seals at the top of
the dewar. In fact, this arrangement rotates so smoothly that it is
possible to perform fixed-field cavity perturbation measurements as
a function of the field orientation, as has recently been
demonstrated for the organic conductor
$\alpha$-(BEDT-TTF)$_2$KHg(SCN)$_4$~\cite{kovalev02}. We generally
use the stepper-motor to control the polar coordinate, while
mechanical control of the cavity end-plate is used to vary the plane
of rotation, i.e. the azimuthal coordinate.

Due to the extremely precise control over both angles, and because
of the need for such precision in recent experiments on
single-molecule magnets which exhibit remarkable sensitivity to the
field orientation~\cite{TakahashiPRB04}, we have found it necessary
to make two modifications to the 3/4" rotating cavity probe for the
purposes of two-axis rotation experiments. The first involves
constraining the cavity assembly and the waveguides within a 3/4"
thin-walled stainless steel tube which is rigidly connected to the
top of the probe. This tube reduces any possible effects caused by
magnetic torque about the probe axis, which could mis-align the
cavity relative to the rotator. The second modification involves
attaching a spindle on the under-side of the end-cap. This spindle
locates into a centering ring attached to the bottom of the PPMS
flow-cryostat, thus preventing the waveguide probe from rotating
off-axis (note that the inner diameter of the cryostat is 1.10" as
opposed to the 3/4" outer diameter of the probe).

\section{Experimental tests}
In this section, we present two examples of unique experiments which
were recently performed using the rotating cavity. We also assess
the performance of the instrument.

\subsection{High-field angle-dependent cyclotron resonance in $\kappa$-(ET)$_2$Cu(NCS)$_2$}
One of the greatest benefits of the rotating cavity is the
possibility to perform angle-dependent electrodynamic studies in the
high-field magnets at the NHMFL. Microwave absorption in a magnetic
field can be used to gain important information about the Fermi
surfaces and other fingerprints associated with the itinerant
electrons in many materials of current fundamental and technological
interest, particularly low-dimensional conductors and
superconductors. These measurements typically require the ability
orient the magnetic field relative to a single-crystal sample. This
can easily be achieved at a low fields by rotating a horizontal
field magnet, changing the currents in orthogonal Helmholtz magnet
pairs, or by rotating the entire measurement apparatus in a split
solenoid with radial access. However, these approaches are currently
limited by the available magnets to fields below about 10~T. While
relatively high-field split-pair magnets are planned in the future
at the NHMFL, these will still be limited to roughly 20~T, since the
split dramatically reduces the attainable field. Thus, an in-situ
sample rotation capability is essential for very high-field studies.
In this section, we highlight these capabilities using as an example
the observation of a novel form of quasi-one-dimensional (Q1D)
cyclotron resonance (CR) in a low-dimensional organic
conductor~\cite{kovalev02}.

The organic charge-transfer-salts (CTSs) $\kappa$-(ET)$_2$X (anion:
X=Cu(NCS)$_2$, I$_3$, Cu[N(CN)$_2$]Br, etc.) are
quasi-two-dimensional (Q2D) conductors consisting of layers of
$\pi$-stacked bis-(ethylenedithio)-tetrathiafulvalene (BEDT-TTF or
ET for short), separated by insulating anion layers. The separation
between the conducting layers is sufficient to treat the ET CTSs as
Q2D systems. At low temperatures, the $\kappa$-(ET)$_2$X CTSs become
superconducting, with many properties similar to the
high-temperature cuprate superconductors, e.g.: an anisotropic
superconducting state~\cite{dressel94:cuncs}, with the possibility
of a nodal energy gap~\cite{izawa01:scgap}; a phase diagram which
consists of an antiferromagnetic insulating phase in close proximity
to a superconducting phase; and an unconventional metallic
phase~\cite{mckenzie97:science}. The FS of the X=Cu(NCS)$_2$ CTS, as
calculated using an extended tight binding model~\cite{ishiguro},
consists of a pair of corrugated Q1D (open) electron sheets and a
closed Q2D hole pocket. When a magnetic field is applied in the
plane of the Q1D FS, charged particles move in response to the
Lorentz force along periodic momentum-space trajectories across the
Q1D FS sheets. During this motion, the Fermi velocity, $v_F$,
remains exactly perpendicular to the FS. Consequently $v_F$ becomes
{\it oscillatory} due to the periodic corrugation of the FS. This
{\it oscillatory} velocity results in a conductivity
resonance$\--$the so-called periodic-orbit-resonance
(POR~\cite{hill97,kovalev02}). This Q1D POR can be distinguished
from the usual CR by its angle dependence~\cite{kovalev02}, which is
given by,

\begin{equation}
{ \frac{\nu}{B_{||}}=\frac{ev_FR_{||}}{h}|\sin(\psi-\psi_\circ)|
 },
 \label{Eqn1}
\end{equation}

\bigskip

\noindent{where $\nu$ is the microwave frequency, $B_{||}$ is the
projection of the magnetic field ($B$) onto the plane of the FS,
$R_{||}$ is the projection of the lattice vector associated with the
FS corrugation onto the FS, and $\psi$ is the angle between the
corrugating axis (along $\psi_\circ$) and $B_{||}$; {\it e} is the
electron charge, and {\it h} is the Plank constant. One can,
therefore, determine {\it $v_F$} and the direction of the
corrugation axis, $\psi_\circ$, by studying angle dependent Q1D
POR.}

Measurements for X=Cu(NCS)$_2$ were carried out using the 33 tesla
resistive magnet at the NHMFL. The high fields are needed in order
to access the metallic phase at low-temperatures ($\sim1.5$~K),
since the superconducting critical field $B_\mathrm{c2}^\perp$ is
$\sim5$~T for fields perpendicular to the layers and
$B_\mathrm{c2}^\parallel \sim 30$~T for fields parallel to the
layers. The flat platelet-shaped sample was mounted on a quartz
pillar, attached to the cavity end-plate [Fig.~4(b)], enabling field
rotations in a plane perpendicular to the highly conducting b-c
plane plane ($\parallel$ to the platelet). Experimental spectra
obtained at different angles are shown in Fig.~6(a).
\begin{figure}
\includegraphics[width=70 mm]{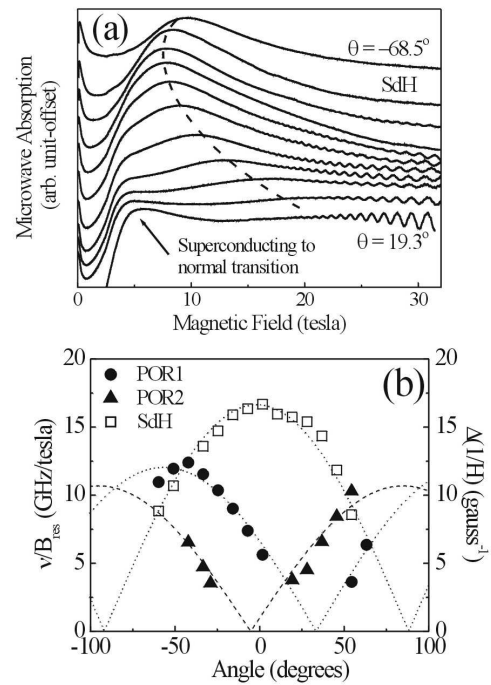}
\caption{\label{fig:CuNCSRaw} (a) 1.4~K experimental traces of the 95~GHz microwave absorption, as a function of the magnetic field strength, for different
field orientations; $\theta$ represents the angle between the least conducting direction ($\perp$ to the b-c plane) and the applied field, and the angle step
is 8.78$^\circ$. The broad absorption peak, which moves to higher fields upon rotating from $-68.5^\circ$ towards $0^\circ$ (indicated by the dashed curve),
corresponds to the Q1D POR. The relatively angle-independent peak at low fields corresponds to a Josephson plasma resonance associated with the superconducting
state~\cite{hill02:jpr}. At high fields, SdH oscillations are observed. These may be used to locate the least conducting direction. (b) A summary of the angle
dependence of two observed series of Q1D POR [only 1 series is shown in (a)], and of the frequency of the SdH oscillations. The dashed curves are theoretical
fits to the data. In the case of the Q1D POR, the fits are used to determine the orientations of the Q1D FS corrugation axes, and the Fermi
velocity~\cite{kovalev02,IJMPB05}.}
\end{figure}
The data contain extremely rich information, including: a Q1D POR, as indicated by the dashed line; Shubnikov-de Haas (SdH) oscillations at high-fields; and a
Josephson plasma resonances (JPR) associated with the superconducting to normal conducting transition at low fields~\cite{hill02:jpr}. The angle dependence of
the POR position [Fig.~6(a)], and the SdH oscillation frequency, are plotted in Fig.~6(b). The SdH oscillations are governed by the closed trajectories on the
hole FS; thus, their angle-dependence should be the same as that for a Q2D CR~\cite{Shoenberg}. However, the angle dependence of the observed POR in Fig.~6 is
clearly quite different from the SdH oscillations. Furthermore, the behavior depends on the plane of rotation (not shown, see ref.~\cite{IJMPB05}). Therefore,
this implies that the POR originates from the Q1D FS. Fitting of the two observed POR series [Fig.~6(b)] to Eq.~1 allows us to determine the orientations of
the corrugation axes at $\psi_\circ=-19.7\pm0.5^\circ$ and $34.7\pm1.5^\circ$ away from the crystallographic $a^*$ axis. These directions correspond to the
{\bf T}$_{10}$ and {\bf T}$_{11}$ directions, where {\bf T}$_{mn}$ is related to the primitive lattice vectors, {\bf T}$_{mn}=m{\bf a}+n{\bf c}$. We also
obtained a value for the Fermi velocity, {\it v$_F$}=3.3 $\times$ 10$^4$ m/s.

\subsection{Angle-dependent high-frequency EPR studies of single-molecule magnets}
\label{smm} Our more recent research efforts have focused on
nanometer scale single-molecule magnets (SMMs) consisting of a core
of strongly exchange-coupled transition metal ions (e.g. Mn, Fe, Ni
or Co) that collectively possess a large magnetic moment per
molecule, thus far up to approximately $51\mu_B$
($S=51/2$)~\cite{Mn25}. SMMs offer a number of advantages over other
types of magnetic nanostructures. Most importantly, they are
monodisperse$\--$each molecule in the crystal has the same spin,
orientation, magnetic anisotropy and molecular
structure~\cite{sessoli03}. They thus enable fundamental studies of
properties intrinsic to magnetic nanostructures that have previously
been inaccessible. For example, recent studies of SMMs have revealed
the quantum nature of the spin-dynamics in a nanomagnet: a
metastable state of the magnetization, say ``spin-up," has been
convincingly shown to decay by quantum tunneling through a magnetic
anisotropy barrier to a ``spin-down" state, in a process called
quantum magnetization tunneling
(QMT)~\cite{sessoli03,Friedman,Thomas}. Remarkably, QMT in SMMs can
be switched on and off, either via a small externally applied field,
or by chemically controlling local exchange interactions between
pairs (dimers) of SMMs, so-called
``exchange-bias"~\cite{hill03:sci}. Thus, the prospects for quantum
control and quantum information processing are very exciting.

All known SMMs possess a dominant uniaxial magnetocrystalline
anisotropy and, to lowest order, their effective spin Hamiltonian
may be written

\
\label{rescond}
\begin{equation}
\hat H = D\hat S_z^2  + g\mu _B \vec{B}.\hat S + \hat H',
\end{equation}
\

\noindent{where $D$ ($< 0$) is the uniaxial anisotropy constant, the
second term represents the Zeeman interaction with an applied field
$B$, and $\hat{H}^\prime$ includes higher order terms in the crystal
field, as well as environmental couplings such as intermolecular
interactions~\cite{hill98:epr,hill03:prl,sessoli03}. This Ising-type
anisotropy is responsible for the energy barrier to magnetization
reversal and the resulting magnetic bistability$\--$factors which
lead to magnetic hysteresis at sufficiently low temperatures. QMT in
zero-field is caused by interactions which lower the symmetry of the
molecule from strictly axial, thereby mixing otherwise degenerate
pure ``spin-up" and ``spin-down" states. QMT rates depend on the
degree of symmetry breaking, which can be determined very precisely
via angle-dependent single-crystal EPR measurements.}

The first SMM, [Mn$_{12}$O$_{12}$(CH$_3$COO)$_{16}$(H$_2$O)$_4$]$\cdot$2CH$_3$- COOH$\cdot$4H$_2$O (Mn$_{12}$-Ac), has become the most widely studied, due to
its giant spin ($S$ = 10) ground state and its high symmetry ($S_4$)~\cite{sessoli03,Lis}. These factors result in the largest known blocking temperature
(T$_B~\sim3$~K) against magnetization relaxation of any SMM. In spite of over 10 years of research, a clear picture has only recently emerged concerning the
symmetry breaking responsible for the QMT in Mn$_{12}$-Ac~\cite{cornia,hill03:prl,delbarco}. Several studies have shown that disorder associated with the
acetic acid solvent in Mn$_{12}$-Ac leads to discrete local environments, resulting in a significant fraction of the molecules ($>50\%$) possessing two-fold
symmetry, with a rhombic crystal-field term $E$ of order 0.01~cm$^{-1}$ (or $E$/$D~\sim 0.02$~\cite{hill03:prl}).

In Fig.~7, we show very high frequency EPR spectra obtained for
several field orientations close to the easy-axis of a
single-crystal sample (volume $\sim0.2$~mm$^3$)~\cite{easy}; the
temperature is 15~K and the frequency is 328~GHz, corresponding to
a very high-order mode of the cavity.
\begin{figure}
\includegraphics[width=70 mm]{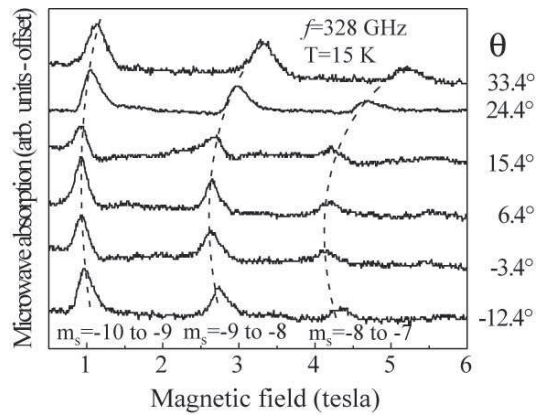}
\caption{\label{fig:highEPR} The angle dependence of the high-frequency EPR spectrum obtained for the Mn$_{12}$-Ac single-molecule magnet~\cite{easy};
$\theta$=0$^\circ$ corresponds to the field applied exactly parallel to the easy-axis of the crystal. The peaks are labeled according the $m_s$ levels involved
in the transitions.}
\end{figure}
The peaks in the transmission through the cavity correspond to EPR, and the resonances have been labeled in the figure according to the $m_s$ levels involved
in the transitions. In spite of the high frequency, the signal-to-noise is good, and the resonances are relatively symmetric, indicating minimal phase and
amplitude mixing. Indeed, the data are comparable in quality with previously published data at much lower frequencies ($<200$~GHz)~\cite{hill02:Fe}. These
experiments were performed using the ESA extension (Gunn diode) with the MVNA, and the sample was mounted on the end-plate. Although the electromagnetic field
distribution at the location of the sample is not known for such a high-order mode of the cavity, it can be assumed that there is a significant $\tilde{H}_1$
component transverse to the applied DC field. Due to the significant axial crystal-field splitting in Mn$_{12}$-Ac (Eq.~2), the ground state ($m_s=-10$) to
first excited state ($m_s=-9$) separation is on the order of 300~GHz in zero field. Therefore, the $m_s=-10\rightarrow-9$ transition is inaccessible below
300~GHz; hence the requirement for high frequencies. At low fields, the positions of the EPR peaks are affected mainly by the field component parallel to the
easy axis of the crystal ($\propto\cos\theta$, where $\theta$ is the angle between the applied field and the easy axis), as observed in Fig.~7. Thus, two-axis
rotation experiments may be used to locate the easy axis.

In Fig.~8, we compare high-resolution angle-dependent high-field
EPR spectra at $\sim51.5$~GHz and 15~K, for field orientations
close to the hard plane ($\theta\sim90^\circ$), obtained for two
forms of Mn$_{12}$: (a) the standard Mn$_{12}$-Ac; and (b) another
high-symmetry ($S_4$) complex,
[Mn$_{12}$O$_{12}$(O$_{2}$CCH$_{2}$Br)$_{16}$(H$_{2}$O)$_{4}]\cdot$4CH$_2$Cl$_2$,
or Mn$_{12}$-BrAc~\cite{petukhov}.
\begin{figure}
\includegraphics[width=62 mm]{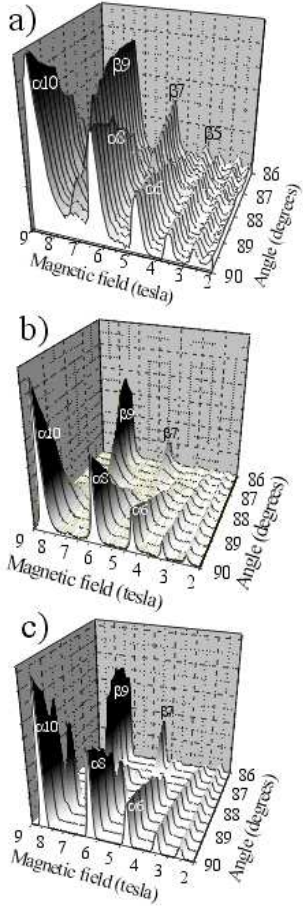}
\caption{\label{fig:mn12data} 3D grey-scale plots of the angle dependence of the low-frequency ($\sim51.5$~GHz, 15~K) hard-plane EPR spectra obtained for (a)
Mn$_{12}$-Ac~\cite{TakahashiPRB04} and (b) Mn$_{12}$-BrAc~\cite{petukhov}; $\theta$=90$^\circ$ corresponds to the field applied exactly parallel to the
hard-plane of the crystal. Simulated spectra (51.5~GHz, 15~K) are displayed for comparison in (c); the simulations were generated using accepted Hamiltonian
parameters, which are essentially identical for the two complexes~\cite{TakahashiPRB04}. See main text and refs.~\cite{TakahashiPRB04,petukhov} for an
explanation of the data.}
\end{figure}
Both data sets were collected in an axial high-field magnet, with the needle-shaped samples mounted directly on the end-plate of the cavity such that
$\tilde{H}_1~\perp~${\bf B}$_\circ$, as illustrated in Fig.~4a. The two Mn$_{12}$ complexes differ only in their ligand and solvent molecules, otherwise the
Mn$_{12}$O$_{12}$ cores are identical. Indeed, the average crystal-field parameters for the two complexes are essentially the
same~\cite{TakahashiPRB04,petukhov}. However, the spectra for Mn$_{12}$-Ac exhibit considerable complexity in comparison to Mn$_{12}$-BrAc, e.g. the EPR peaks
are broader, asymmetric, and contain several fine structures (see ref.~\cite{TakahashiPRB04} for more details). Perhaps the most striking difference in the
spectra for the two complexes is the significant overlap of the $\alpha$ and $\beta$ series of resonances for the Mn$_{12}$-Ac sample. The origin of the
$\alpha$ and $\beta$ resonances is discussed in detail in ref.~\cite{TakahashiPRB04}, and is beyond the scope of this article. Nevertheless, simulated spectra
(also for 51~GHz and 15~K) are displayed in Fig.~8c, which indicate that the $\alpha10$ and $\alpha8$ peaks should {\em not} overlap in angle with the $\beta9$
resonance. In this respect, the data for the Mn$_{12}$-BrAc complex exhibit good agreement with the simulations. In contrast, the overlapping of the $\alpha$
and $\beta$ resonances have provided the first indications that the molecular easy axes for the Mn$_{12}$-Ac complex are tilted locally (up to
$1.7^\circ$~\cite{TakahashiPRB04}). Such tilting can be understood within the context of the solvent disorder picture~\cite{cornia}: the various fine
structures, and angular overlap of different portions of the EPR spectra, reflect a (discrete) distribution of local environments caused by the solvent
disorder. Using the two axis rotation capabilities, more recent studies have confirmed the discrete nature of the local environments, e.g. the molecular easy
axis tilts are confined to orthogonal planes, and each of the EPR fine structures has its own distinct angle dependence~\cite{HillPoly05}. Meanwhile, none of
the fine-structures are seen in the Mn$_{12}$-BrAc complex, thus confirming the idea that the acetic acid solvent significantly influences the magnetization
dynamics in Mn$_{12}$-Ac. Indeed, the EPR spectra for Mn$_{12}$-BrAc provide unprecedented resolution, allowing for unique spectroscopic insights into
high-symmetry giant spin SMMs~\cite{petukhov}.



\begin{acknowledgments}
The authors appreciate useful discussions with Rachel Edwards,
Alexey Kovalev and Bruce Brandt. We would also like to thank Daniel
Benjamin, Marc Link, Mike Herlevich and John Van Leer for technical
assistance. This work was supported by the National Science
Foundation (DMR0196461 and DMR0239481), and by Research Corporation.
Work at the NHMFL is supported by a cooperative agreement between
the State of Florida and the National Science Foundation through
NSF-DMR-0084173.
\end{acknowledgments}


\begin{thebibliography}{58}
\bibitem[{not()}]{note1}
\bibinfo{note}{The larger worm gear used in the 1st generation cavities was
  commercially manufactured from teflon.}
\end{thebibliography}
\end{document}